\documentclass[aps,onecolumn,superscriptaddress]{revtex4}

\usepackage{amsmath,amssymb,graphicx}
\usepackage{color}

\definecolor{yblue}{rgb}{0.06, 0.3, 0.57}
\usepackage{hyperref}
\hypersetup{colorlinks=true,linkcolor=yblue,citecolor=yblue,urlcolor=yblue}

\graphicspath{{fig-ps/}}

\begin{document}

\title{Instabilities via Negative Krein Signature in a
  Non-Conservative DNLS Model}

\author{P. G. Kevrekidis}
%\email{kevrekid@math.umass.edu}
\affiliation{Department of Mathematics and Statistics, University of Massachusetts,
Amherst, Massachusetts 01003-4515 USA}

\begin{abstract}
  In the present work we consider a model that has been proposed
  at the  continuum level for self-defocusing
  nonlinearities in atomic BECs in order to
  capture phenomenologically the loss of condensate atoms to thermal
  ones. We explore a model
  combining dispersion, nonlinearity and gain/loss at the discrete level,
  %using this 
  %prototypical setting  to
  illustrate the idea that modes associated with negative ``energy''
  (mathematically: negative Krein signature) can give rise to instability
  of excited states when gain/loss terms are introduced in a nonlinear
  dynamical lattice.
  We showcase this idea by considering one-, two- and three-site discrete
  modes, exploring their stability
  via analytical approximations, and corroborating
  their continuation over the gain/loss parameter numerically,
  as well as manifesting through direct numerical simulations their
  unstable nonlinear dynamics.
\end{abstract}

\maketitle

\section{Introduction}

One of the principal nonlinear lattice dynamical models that
are used to examine the evolution of coherent structures in
discrete systems is the discrete nonlinear Schr{\"o}dinger
equation (DNLS)~\cite{dnlsbook}. One of the main reasons for its
popularity is that it contains the prototypical
ingredients for the manifestation of interesting phenomena,
namely lattice (discrete) dispersion and nonlinearity. These
features arise (in this very combination) in a wide variety
of physical contexts. Most notable among them in the past
two decades have, arguably,
been the study of optical waveguide arrays~\cite{dnc,moti},
as well as the evolution of atomic Bose-Einstein condensates (BECs) in the
realm of optical lattice potentials~\cite{ober}.
These settings have, in turn, enabled not only the theoretical
study, but also importantly the experimental observation of
a wide array of features. Among them, a partial list includes
discrete diffraction~\cite{yaron} and diffraction management~\cite{yaron1},
lattice solitons~\cite{yaron,yaron2} and vortices~\cite{neshev,fleischer}, 
Talbot revivals~\cite{christo2}, and $\mathcal{PT}$-symmetry
breaking~\cite{kip}, among many others. 

On the other hand, in the context of atomic BECs, as the comprehension
of their mean-field features~\cite{stringari,siambook}
is becoming more mature,
it is natural to seek a deeper understanding of deviations from the
standard setting, such as those involving thermal~\cite{proukbook} and
quantum~\cite{ldcbook}
fluctuations. For the former, a simple phenomenological model
of interest was introduced by Pitaevskii~\cite{d27} and was
later shown to be relevant from a microscopic perspective~\cite{d28}.
In fact, this so-called dissipative Gross-Pitaevskii equation (DGPE)
model was also favorably compared~\cite{d17},
as regards nonlinear wave dynamics,
with more elaborate models such as the so-called stochastic Gross-Pitaevskii
equation (SGPE) model; see, e.g.,~\cite{d29} for a review of the latter.
In addition, the DGPE model was used to predict the anti-damped
motion of pairs (and triplets) of dark solitons~\cite{d30} and
also of multi-component solitons~\cite{ddb}, as well as to quantify
the spiraling out of the condensate of vortices~\cite{nick}, that
was recently observed  experimentally~\cite{korean}.

Our aim in the present work is to combine these two notions,
i.e., to consider a discrete variant of the DGPE model.
To simplify (technical) matters, we explore the case of the focusing
nonlinearity [which corresponds to
  attractive interactions for a condensate inside an optical lattice],
however the results can 
be extended to the defocusing 
case; see, e.g., for one such example~\cite{gaphadi}.
%~\footnote{It is
%  relevant to highlight here that in this case the model is
%  not purely dissipative. For this reason, we will refer to it
%  as a model bearing gain and loss. Further explanation of these
%  notions is given in the first paragraph of the Conclusions.}. 
We consider the relevant model close to the so-called anti-continuum
limit, of near-vanishing coupling and explore the role of the
gain/loss parameter (that we denote by $\gamma$) in the stability
properties of some of the principal discrete soliton solutions.
{It is important to highlight that the realistic
values of this parameter are quite small (e.g., the work
of~\cite{nick} estimates them as being between $0.00023$ and $0.0023$
for temperatures between $10$ and $100$nK in atomic BECs),
nevertheless their stability impact is quite significant.}
The main
contribution of our work is that it enables the generalization of
the stability considerations (based on Lyapunov-Schmidt reductions
and associated spectral calculations~\cite{pkf05}) to this
setting involving the presence of gain/loss. The main findings that
we obtain are two-fold: on the one hand, the eigenvalues that are
found to possess negative ``energy'' (so-called negative Krein
signature, mathematically) bifurcate in {\it the opposite direction} than
the rest of the spectrum when gain/loss is introduced. This is in line
with the general theory of~\cite{kks04}, but here the detailed
calculation of these eigenvalues is provided and corroborated by
our numerical findings. On the other hand, we observe that also
the two-fold symmetric eigenvalue at the origin {\it also splits}
and one of the two members of the pair moves along the real axis, leading
to an instability of the most fundamental single-site soliton
state. Once again, this can be attributed
to the presence of gain/loss in the model.

This model can be naturally placed in the context of discrete variants
of the complex Ginzburg-Landau (GL) equation (for a review of its continuum
form, see e.g.~\cite{aranson}). In what follows, we give some prototypical
examples of relevant works in this GL direction from a discrete perspective
and subsequently highlight the crucial elements of novelty of the present
contribution. In~\cite{efremidis1}, some important families of solutions in
the cubic-quintic discrete GL equation were identified, including rather
exotic ones (such as cusp-like solitons) and their bifurcation diagrams
were constructed. In the follow up work of~\cite{efremidis2}, generalizations
of on-site and inter-site solutions (with one-, and two or four-peaks,
respectively) of different types were considered in the context of
two spatial dimensions. More mathematically inclined works such as
that of~\cite{karachalios} were concerned with the potential convergence
of the solutions of these discrete GL equations to global attractors
(or their potential blowup). Another example where such models were
considered was that of~\cite{kiselev} where a discrete model bearing
a saturable nonlinearity was put forth and both plane wave solutions,
as well as discrete solitons and their stability were examined.
It is important to highlight here that while numerous important works
exist in the subject (of which the above references
constitute only a representative sample), it is our understanding that
there is no earlier setting putting forth an analytical evaluation
of the relevant stability eigenvalues, leading to a straightforward/
easy-to-use expression such as that of Eqs. (11)-(12) below.
Alongside this result, emerges an important interpretation about
the expectation of stability of different modes. In particular,
an explicit demonstration arises of the fact that the continuous
spectrum in this case will move in a direction opposite to that
of the so-called negative energy eigenvalues. To the best of
our knowledge, such eigenvalue expressions, such stability intuition
and its quantitative corroboration with numerical computation
(in both stability and dynamics) is unprecedented in the discrete
GL literature.

Our presentation of the above findings will be structured as follows.
First, in section II,
we will present the model and the associated general theory
for its spectral properties.
Then, in section III, we will provide existence, stability
and dynamical considerations that corroborate our analytical
theory for the cases of one-, two-
and three-site solutions of the model.
More specifically, we will examine the stability of
the solutions for different values of $\gamma$. Having obtained
that, we will move on to the consideration of the dynamics
of the different  instabilities in select case examples. We will explain
how the direct dynamical results reflect the conclusions of
our stability analysis. Finally, in section IV, we will summarize
our findings and provide a number of directions for future
study.

\section{Theoretical Analysis}

Our starting point will be the discrete form of the DGPE
as:
\begin{eqnarray}
  (i -\gamma) \dot{u}_n=-\epsilon \left(u_{n+1} + u_{n-1}\right)
  - |u_n|^2 u_n + \mu u_n.
  \label{dg1}
\end{eqnarray}
%From the expression, one immediately realizes that our model
%is a combination of the standard DGPE (see, e.g.,~\cite{d17})
%and the DNLS~\cite{dnlsbook} models.
Here, $\gamma>0$ plays
the role of the gain/loss parameter, while $\epsilon$ controls
the coupling between adjacent nodes of our 1d lattice.
In detailed calculations that will follow including in the next section,
the frequency parameter $\mu$ will be selected as $\mu=1+2 \epsilon$
(for convenience, given its tunability).

The steady state problem is the same in this case as it is
for the standard DNLS, in particular for a given
$\epsilon$, the same stationary states that exist
for $\gamma=0$ also persist for $\gamma$ finite. Hence,
$\gamma$ does not affect the existence properties which
are well-known; see, e.g., the review
in Chapter 2 of~\cite{dnlsbook}. It is long established,
in particular, that although at the anti-continuum (AC)~\cite{MA94}
limit of $\epsilon=0$, solutions of
arbitrary phase $u_n=e^{i \theta_n}$ are available, for
finite $\epsilon$, the Lyapunov-Schmidt condition
\begin{eqnarray}
  \sin(\theta_{n+1}-\theta_n) + \sin(\theta_{n-1}-\theta_n)=0
  \label{dg2}
\end{eqnarray}
needs to be enforced. Coupled to the stipulated asymptotic decay of the
solution at $n \rightarrow \pm \infty$, this leads to the
phases of the solution being ``locked'' to $\theta_n=0, \pi$~\cite{pkf05}.
Solutions involving a single
site (the ground state), 2-sites and 3-sites are often
examined; cf., e.g.,~\cite{pkf05}.

The main element where the gain/loss parameter plays a role,
however, is the stability of the solution. We now assume
that the (real) solution $u_n^{0}$ is the state around which we
linearize, in order to explore the spectral response of
the waveform.
The latter is the principal contribution of the present
work in the considered discrete GL setting.
Then the linearization ansatz will read:
\begin{eqnarray}
  u_n =u_n^0 + \delta (A_n + i B_n),
  \label{dg3}
\end{eqnarray}
from which we extract the equations to O$(\delta)$ for the
resulting dynamical system, in order to examine the spectrum
of small perturbations. We  thus find:
\begin{eqnarray}
\left( \begin{array}{cc}
-\gamma & -1 \\
1  & -\gamma \\
    \end{array} \right)
\left( \begin{array}{c}
\dot{A}_n \\
\dot{B}_n \\
    \end{array} \right)= 
\left( \begin{array}{cc}
{\cal L}_+ & 0 \\
0  & {\cal L}_- \\
\end{array} \right)
\left( \begin{array}{c}
{A}_n \\
{B}_n \\
\end{array} \right).
\label{dg4}
\end{eqnarray}
Here, we have defined
\begin{eqnarray}
  {\cal L}_+ &=&-\epsilon (\Delta_2 + 2) + \mu-3 (u_n^0)^2
  \label{dg4a}
  \\
    {\cal L}_- &=&-\epsilon (\Delta_2 + 2) + \mu- (u_n^0)^2,
    \label{dg4b}
\end{eqnarray}
where $\Delta_2$ symbolizes the discrete Laplacian with unit
spacing $\Delta_2 u_n=u_{n+1}+u_{n-1}-2 u_n$.

The resulting equations, upon decomposing $A_n=a_n e^{\lambda t}$
and $B_n=b_n e^{\lambda t}$ are
\begin{eqnarray}
  \left( \tilde{\lambda} + \gamma {\cal L}_+ \right) a_n &=& {\cal L}_- b_n
  \label{dg5}
  \\
  \left( \tilde{\lambda} + \gamma {\cal L}_- \right) b_n &=& - {\cal L}_+ a_n.
  \label{dg6}
\end{eqnarray}
Notice also that due to the inversion of the matrix containing
the $\gamma$'s (and the associated determinant), in all the expressions
below we rename ${\lambda} (\gamma^2+1)= \tilde{\lambda}$.
In the $\gamma \rightarrow 0$ limit, $\tilde{\lambda}$
coincides with $\lambda$. 
%and so in order to retrieve
%the ``true'' eigenvalues of the problem, from the $\lambda$ obtained below
%we need to divide the associated expression by $\gamma^2+1$.

Assuming now that ${\cal L}_+$ is invertible, which it
is close to the AC limit, where it becomes a multiplicative operator,
%$(\lambda+\gamma {\cal L}_+)$ is invertible (see also
%below),
we can write:
\begin{eqnarray}
  \left(\tilde{\lambda} + \gamma  {\cal L}_+ \right) {\cal L}_+^{-1}
  \left(\tilde{\lambda} + \gamma  {\cal L}_- \right) b_n=-{\cal L}_- b_n
  %\left(\lambda + \gamma  {\cal L}_- \right) b_n = - {\cal L}_+
 % \left(\lambda+\gamma {\cal L}_+\right)^{-1} {\cal L}_- b_n.
  \label{dg7}
\end{eqnarray}
Now, we take advantage of the fact that near the AC limit
$(u_n^0)^2=1$ for excited sites (given our selection of $\mu$),
hence the operator
${\cal L}_+$ is a multiplicative one by $-2$ (and thus its
inverse also a multiplicative operator by $-1/2$). Then,
the expression of Eq.~(\ref{dg7}) becomes
\begin{eqnarray}
   \left( \tilde{\lambda} -2 \gamma \right) \left( \tilde{\lambda} + \gamma s \right)=
2 s
  %\frac{2}{\lambda -2 \gamma} s,
  \label{dg8}
\end{eqnarray}
where $s=(b_n, {\cal L}_- b_n)$ denote the small eigenvalues of
the operator ${\cal L}_-$ that have been previously computed
explicitly in the Hamiltonian limit;
see e.g.,~\cite{pkf05}. This then leads to the
principal result of the present work, namely that the model with
gain/loss will possess eigenvalues of the form:
\begin{eqnarray}
  \tilde{\lambda}=\frac{\gamma (2-s) \pm \sqrt{\gamma^2 (s-2)^2 +8 s
  (\gamma^2+1)}}{2}.
  \label{dg9}
\end{eqnarray}
    {We will bear in mind in what follows (also, per the
      motivating example mentioned in the Introduction) that
      practically $\gamma^2 \ll 1$.}
We now consider some special cases. We first mention
the one in the immediate vicinity
of the AC limit, whereby $s={\rm O}(\epsilon)  \ll 1$. Then,
\begin{eqnarray}
  \tilde{\lambda}= \gamma \pm \sqrt{\gamma^2 + 2 s}.
  \label{dg10}
\end{eqnarray}
I.e., for the eigenvalues that bifurcate from $0$, of which, as
the theory of~\cite{pkf05} suggests, there are at least
$N-1$ pairs in the Hamiltonian case (we will treat the eigenvalue
pair at $0$ separately below), we find that they
will {\it generically} lead to instability; $N$ here
denotes the number of excited nodes/sites of the lattice.
This is because if
$s>0$ (i.e., for eigenvalues leading to real instabilities even
in the Hamiltonian case), at least one of them will be
real and positive even in the presence of the terms involving
$\gamma$, per Eq.~(\ref{dg10}).
On the other hand, if $s<0$, as occurs for the eigenvalues
of negative energy in the Hamiltonian case of~\cite{pkf05}, we have
the following. For sufficiently small $\gamma$, i.e.,
$\gamma < \sqrt{2 |s|}$, the relevant imaginary (for the Hamiltonian
case) eigenvalues will turn complex with a positive real
part. Eventually, these eigenvalues will collide
on the positive real axis at $\gamma=\sqrt{2 |s|}$, splitting off as real
thereafter.
%$\gamma < \sqrt{2 s}$ (and $s>0$; if $s<0$, then the eigenvalues
%are already unstable for $\gamma=0$), and these eigenvalues will be complex,
%while if $\gamma > \sqrt{2 s}$, they will be real with at least
%one of the two being larger than $0$, hence associated with an
%instability. The critical point which in this case of $s \ll 1$
%occurs at  $\gamma = \sqrt{2 s}$ signals the collision of the
%two complex eigenvalues on the positive real axis, as a result
%of which they form a real pair.
It is particularly relevant
to point out that {\it all} the rest of the eigenvalues
move to the left half spectral plane (stable eigendirections),
reflecting the lossiness present in the
problem. Nevertheless, these eigenvalues that emerged from the origin
in the Hamiltonian case of~\cite{pkf05},
because of their negative Krein signature (the sign of $s$ yields
the Krein signature of these eigenvalues; cf. Eq. (3.16) in~\cite{pkf05}),
they move in the {\it opposite} direction, as our main result
clearly indicates and give rise to pairs of complex eigenvalues.
If there are $N-1$ such eigenvalues with $s<0$ for an $N$-site configuration
(as is the case when adjacent excited sites have alternating phases),
then there are going to be $N-1$ complex pairs of associated
eigenvalues emerging for $\gamma>0$.
%from the inclusion of gain/loss.

A special mention deserves to be made for the limit of $s=0$.
It turns out that the above formulae yield the correct result for
the asymptotic case with $s=0$. There exists, as is well-known~\cite{dnlsbook},
a pair of eigenvalues at the origin, due to the phase invariance
in the Hamiltonian case of $\gamma=0$.
The solution itself remains an eigenvector with vanishing eigenvalue in
this case, however the presence of gain/loss avoids the persistence of the
generalized corresponding eigenmode. It is for that reason that the
eigenvalues (of this pair) in this context split when
$\gamma \neq 0$, becoming $\tilde{\lambda}=2 \gamma$ and $0$,
according to Eq.~(\ref{dg10}). We  will see indeed that the relevant expression
provides a very good approximation to the corresponding unstable mode in what
follows.

Lastly, it is important to point out that it is possible to
follow the same type
of analysis for the continuous spectrum eigenmodes, taking into
consideration the facts that:

(a) the relevant eigenvalues can be identified via linearization around
the vanishing state with $u_n^0=0$;

(b) in this limit ${\cal L}_+={\cal L}_-$.

Then, the relevant calculation stemming from Eq.~(\ref{dg7}) for
plane waves $\propto e^{i (k n - \omega t)}$ is considerably simplified
and gives rise to a continuous spectrum explicitly calculable as:
\begin{eqnarray}
  \tilde{\lambda}=\left(1+ 4 \epsilon \sin^2(\frac{k}{2})\right) (-\gamma \pm i).
  \label{dg11}
\end{eqnarray}
It is straightforward to see that the introduction of the gain/loss
parameter $\gamma$ sends the continuous spectrum to the  left half
plane, contrary --as we saw above-- to what is the case for the eigenvalues
of negative signature.

%{\it Introduction.} 
%%%%%%%%%%%%%%%%%%%%%%%%%%%%%%%%%%%%%%%%%%%%%%%%%%%%%%%%%%%%%%%%%%%%%%%%%
%\begin{figure*}[thb]
%\includegraphics[width=17.55cm]{dynamicsnew_densTN.jpg.ps}\\[-0.4ex]
%\includegraphics[width=17.55cm]{dynamicsnew_phasT.jpg.ps}\\[ 1.0ex]
%\includegraphics[width=17.55cm]{dynamics_densTN.jpg.ps}   \\[-0.4ex]
%\includegraphics[width=17.55cm]{dynamics_phasT.jpg.ps}
%\caption{
%
%}
%\label{Dynamics}
%\end{figure*}
%%%%%%%%%%%%%%%%%%%%%%%%%%%%%%%%%%%%%%%%%%%%%%%%%%%%%%%%%%%%%%%%%%%%%%%%%

\section{Numerical Computations}

We now explore a diverse array of states that are well-known
to exist in the DNLS model, since the steady state problem,
being the same as in the case of DNLS still gives rise to such
states. In what follows, we will set $\epsilon=0.01$ in the vicinity
of the AC limit and will vary $\gamma$ as a parameter of the model.

\begin{figure}
\includegraphics[height=4.5cm]{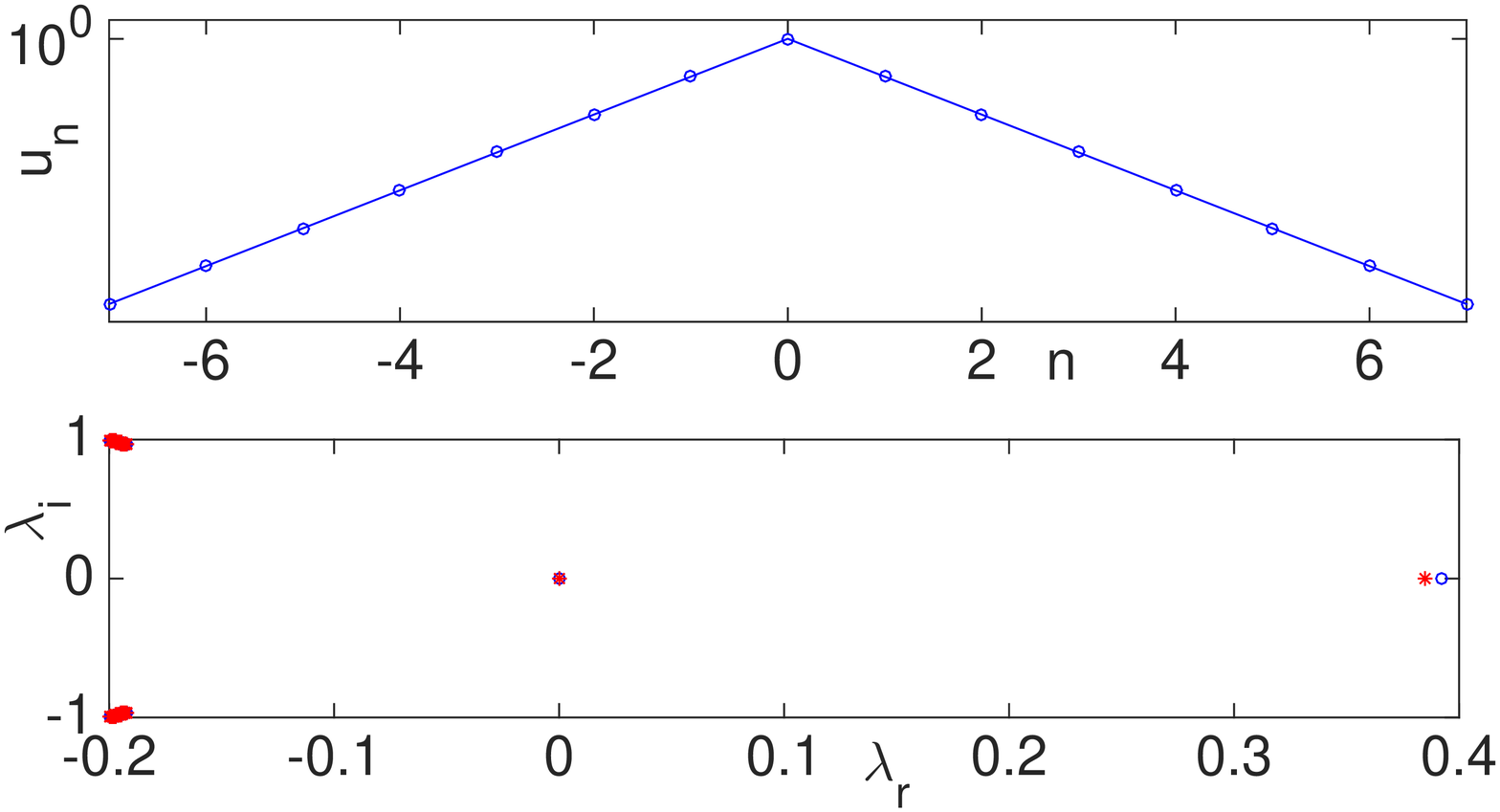}
\includegraphics[height=4.5cm]{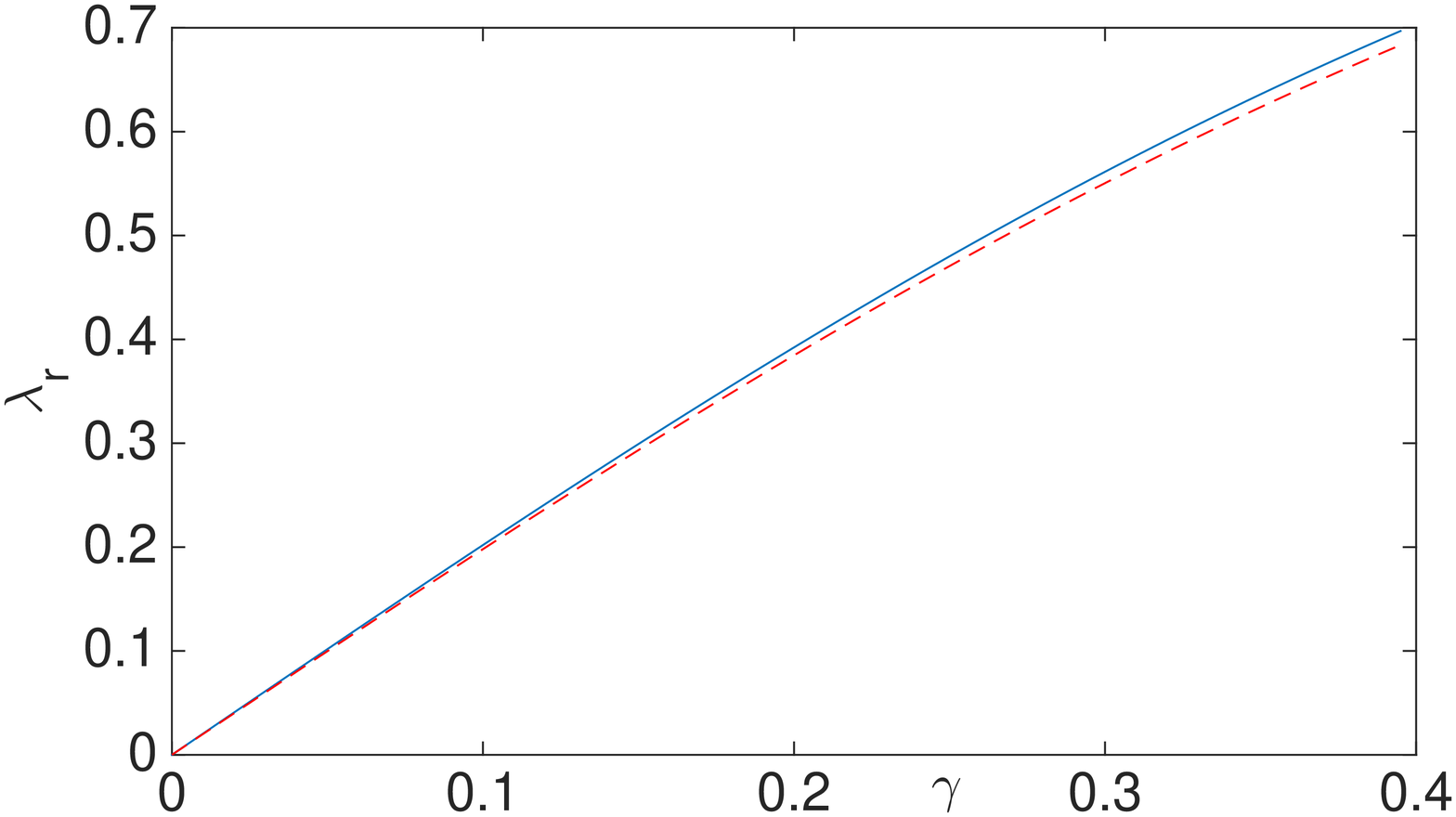}
\caption{The left panel illustrates the spatial profile in
  a semi-log plot, indicating the exponential decay, as
  well as the corresponding spectral plane for $\gamma=0.2$.
  In the latter, both the numerically identified spectral modes
  $\lambda=\lambda_r + i \lambda_i$
  are given (as blue circles) and the theoretical predictions
  stemming from application of Eq.~(\ref{dg7}) for the different
  modes are provided (as red stars) for comparison. The right panel
  shows the unstable mode exiting as real from $(0,0)$ (solid blue
  line) and its corresponding theoretical prediction ($2 \gamma/(\gamma^2+1)$
  as red dashed line).}
\label{figd1}
\end{figure}

We start from the single-site configuration of Fig.~\ref{figd1}.
The left panel of the figure illustrates the exponential decay
of the spatial profile of the configuration.
The spectrum of this mode for $\gamma=0$ (the ground
state of the Hamiltonian problem) features a pair of eigenvalues
at the origin, as well as a continuous spectrum extending
across the interval $\pm i (1 + 4 \epsilon \sin^2(k/2))$.
As soon as $\gamma>0$ is introduced, the
latter modes, move to the left half plane as is illustrated in
the bottom panel of Fig.~\ref{figd1}. On the other hand,
in line with the predictions of Eqs.~(\ref{dg9})-(\ref{dg10}),
the pair of modes at $0$ splits; one of them remains at the
origin, due to the phase invariance, while the other moves
along the real line with $\lambda=2 \gamma/(\gamma^2+1)$
and gives rise to the instability of the solution.
The comparison of the theoretical prediction with the
numerical result is excellent
(for this eigenmode, as well as for the entire spectrum)
and is shown in the right panel
of Fig.~\ref{figd1}.

%%%%%%%%%%%%%%%%%%%%%%%%%%%%%%%%%%%%%%%%%%%%%%%%%%%%%%%%%%%%%%%%%%%%%%%%%
%%%%%%%%%%%%%%%%%%%%%%%%%%%%%%%%%%%%%%%%%%%%%%%%%%%%%%%%%%%%%%%%%%%%%%%%%

The next case that we consider is that of the so-called anti-symmetric
or twisted localized modes~\cite{darmanyan}. Such modes are well-known
to be stable in the Hamiltonian case of
$\gamma=0$ near the AC limit, due to a pair at the origin (phase
invariance) and another pair $\lambda= \pm 2 \sqrt{\epsilon} i$,
associated with $s=-2 \epsilon$~\cite{pkf05}. This gives an immediate
prediction that can be used, in addition to the predictions
that we had before for the mode emerging from $(0,0)$ and the continuous
spectrum in the spectral plane $(\lambda_r,\lambda_i)$ of eigenvalues
$\lambda= \lambda_r + i \lambda_i$. In this case, in the left panel
of the figure we show two distinct cases, namely $\gamma=0.1$ and
$0.3$. This is because the two of them are separated by the critical
point $\gamma=\sqrt{2 |s|}$ of Eq.~(\ref{dg10}). As predicted in our theory of
the previous section, the modes become immediately unstable (for
$\gamma > 0$), giving
rise to a complex pair. Eventually, however, the two eigenvalues
of the pair collide at $\gamma=\sqrt{2 |s|}=2 \sqrt{\epsilon}$ ($=0.2$
in this case) and subsequently exit along the real axis. We find
this feature to be in excellent agreement once again with the
numerical findings both as regards the general phenomenology, but
also as regards the quantitative aspects (critical point, etc.).

%%%%%%%%%%%%%%%%%%%%%%%%%%%%%%%%%%%%%%%%%%%%%%%%%%%%%%%%%%%%%%%%%%%%%%%%%
\begin{figure}
\includegraphics[height=3.5cm,width=5cm]{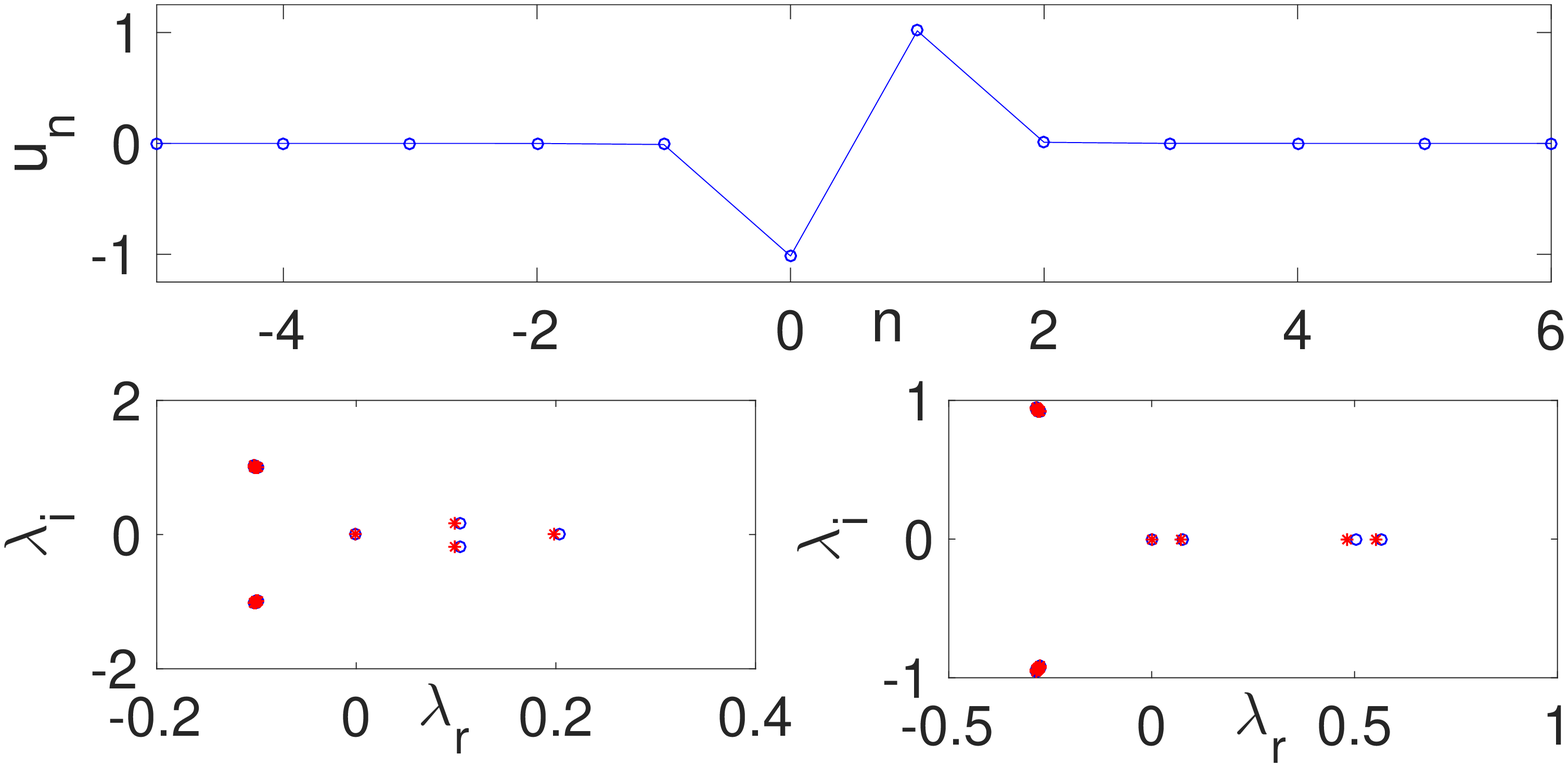}
\includegraphics[height=3.5cm,width=5cm]{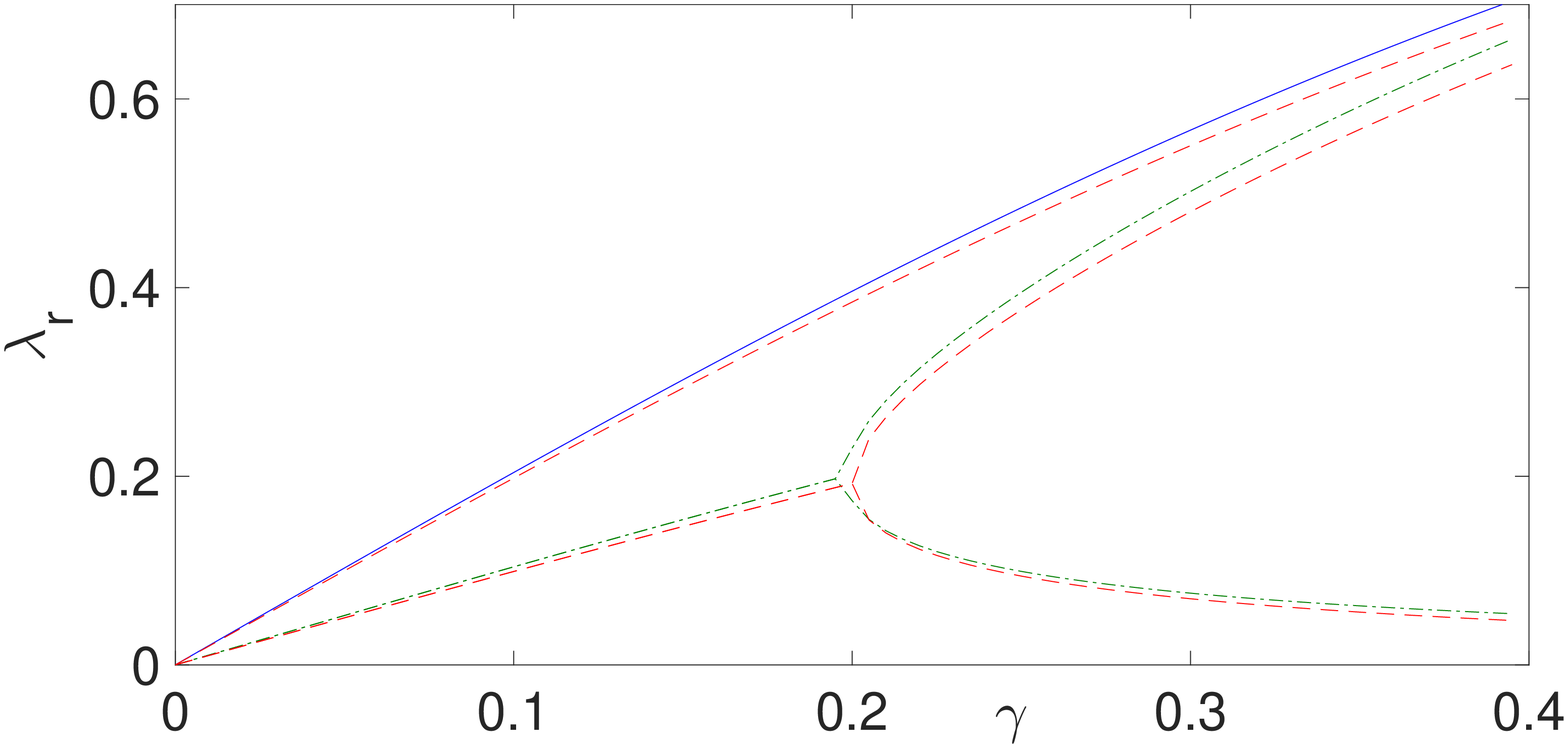}
\includegraphics[height=3.5cm,width=5cm]{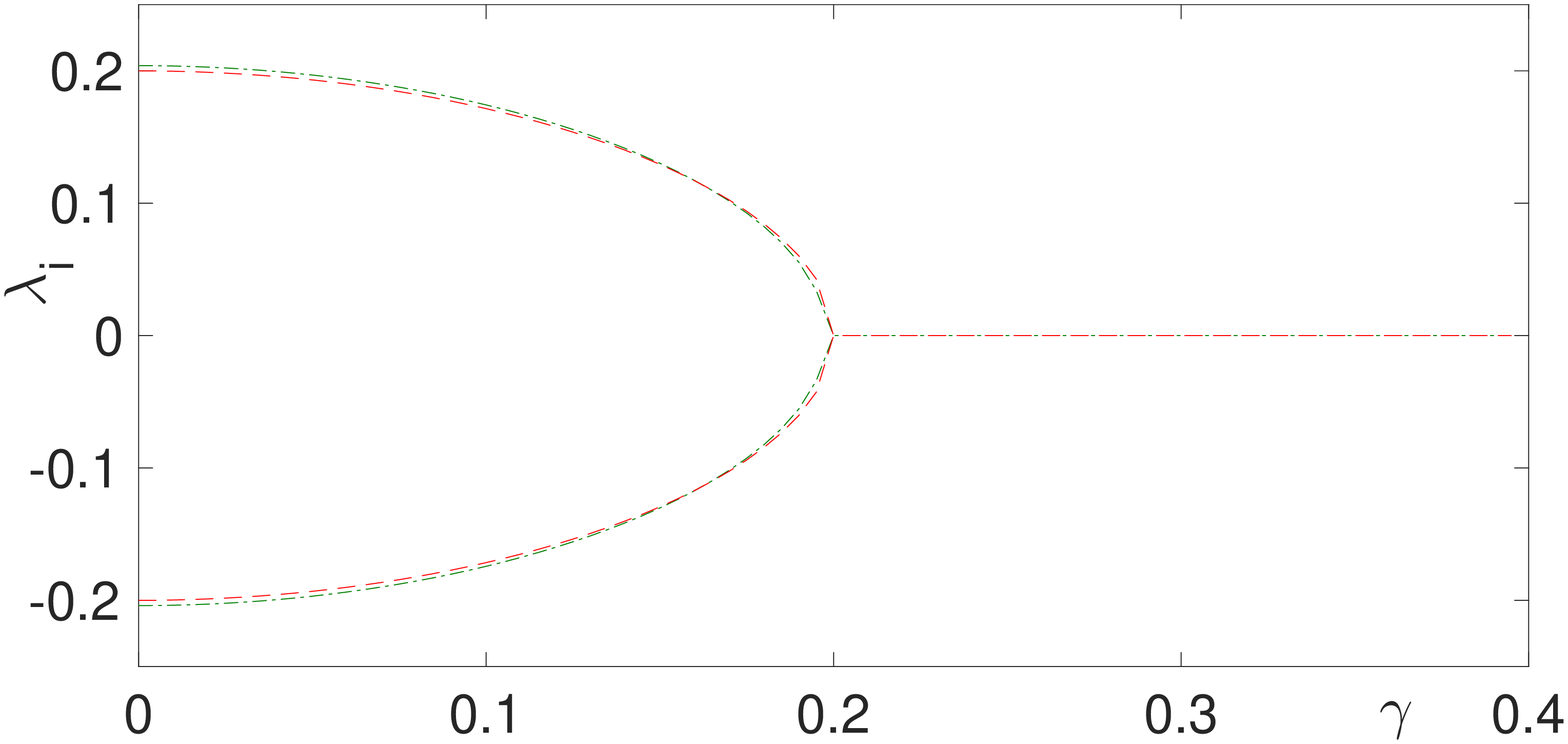}
\caption{The figure considers the case of the two-site,
  out-of-phase mode (sometimes referred to as ``twisted'' localized
  mode~\cite{darmanyan}). The left panel illustrates
  once again the stationary state, while its corresponding
  spectrum is shown for $\gamma=0.1$ and $0.3$ (in excellent
  comparison between the
  blue circles of the numerics and the red stars of the theory).
  The middle and right panels illustrate the dominant (unstable)
  eigenmodes and how they move: one of them, as before and using
  also the same symbolism, emerges
  from $(0,0)$. However, importantly, two more become
  immediately complex (see the green dash-dotted lines),
  in excellent agreement with the theoretical prediction
  (red dashed lines) and
  eventually collide near $\gamma=2 \sqrt{\epsilon}=0.2$,
  giving rise to a real pair thereafter. }
\label{figd2}
\end{figure}
%%%%%%%%%%%%%%%%%%%%%%%%%%%%%%%%%%%%%%%%%%%%%%%%%%%%%%%%%%%%%%%%%%%%%%%%%

We now turn to the case of the two-site in-phase configuration.
This is a state that is well-known in the Hamiltonian limit
to be unstable for all values of the coupling $\epsilon$, due
to a real pair $\lambda=\pm \sqrt{2 s}=\pm 2 \sqrt{\epsilon}$.
For $\gamma > 0$, using $s=2 \epsilon$,
we find that one of the members of the real pair (the positive
one) grows rapidly in excellent agreement with the predictions
of Eqs.~(\ref{dg9})-(\ref{dg10}), while the negative counterpart
tends to 0. Additionally, the mode from $(0,0)$ is still present
(although now the instability is dominated by the already real
mode at $\gamma=0$). Finally, the continuous spectrum moves again
to the left half plane. All the modes are captured very accurately by
the theoretical analysis.

%%%%%%%%%%%%%%%%%%%%%%%%%%%%%%%%%%%%%%%%%%%%%%%%%%%%%%%%%%%%%%%%%%%%%%%%%
\begin{figure}
\includegraphics[height=3.5cm,width=5cm]{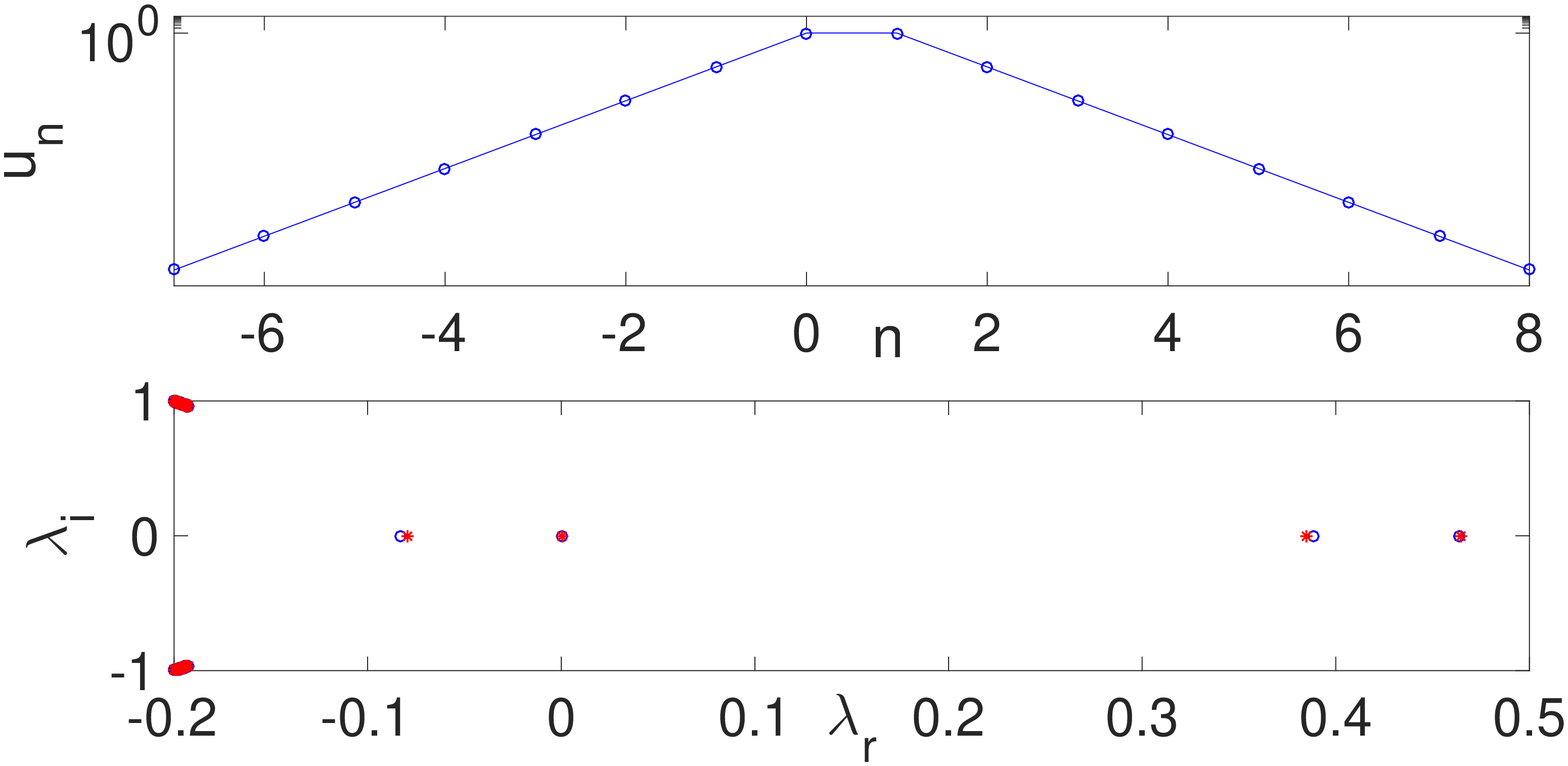}
\includegraphics[height=3.5cm,width=5cm]{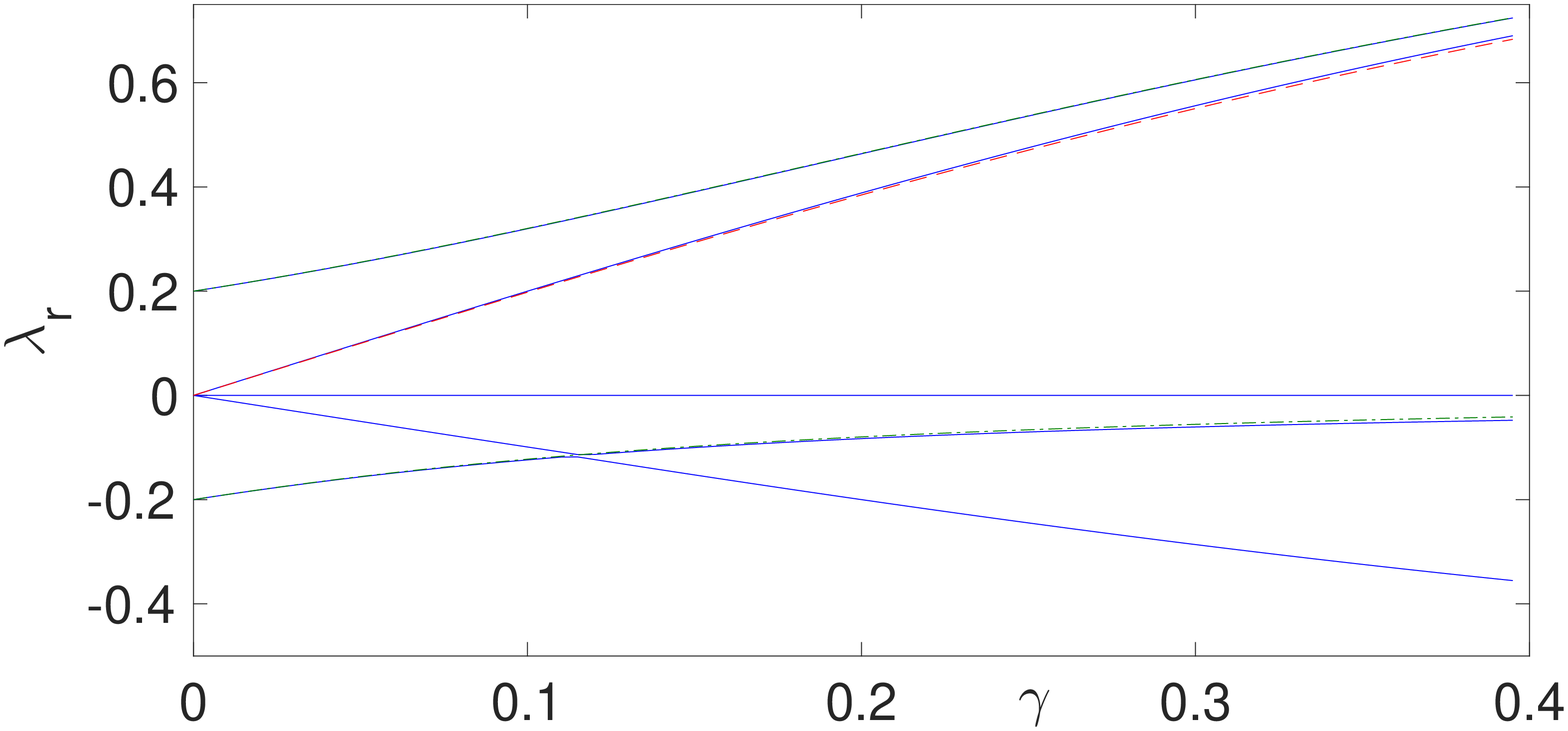}
\caption{This figure concerns the in-phase mode with two
  excited sites. The mode also decays exponentially and is unstable
  already  from the limit of $\gamma=0$ and onwards. The spectral
  plane of the case with $\gamma=0.2$ showcases the excellent agreement
  with the theoretical results. The right panel yields all the numerical
  modes (the unstable pair of $\gamma=0$ --starting at $\pm 0.2$ and
  growing--, the eigenvalue growing from the origin, and the
  continuous spectrum moving to the left i.e., downward in the panel)
  in solid blue line. The red-dashed is the theory for the mode
  emerging from $(0,0)$, and the green dash-dotted curves that can barely
  be distinguished from the blue solid ones showcase the excellent
  agreement of the modes predicted by Eqs.~(\ref{dg9})-(\ref{dg10})
  for $s=2 \epsilon$.}
\label{figd3}
\end{figure}
%%%%%%%%%%%%%%%%%%%%%%%%%%%%%%%%%%%%%%%%%%%%%%%%%%%%%%%%%%%%%%%%%%%%%%%%%

\begin{figure}
\includegraphics[height=3.5cm,width=5cm]{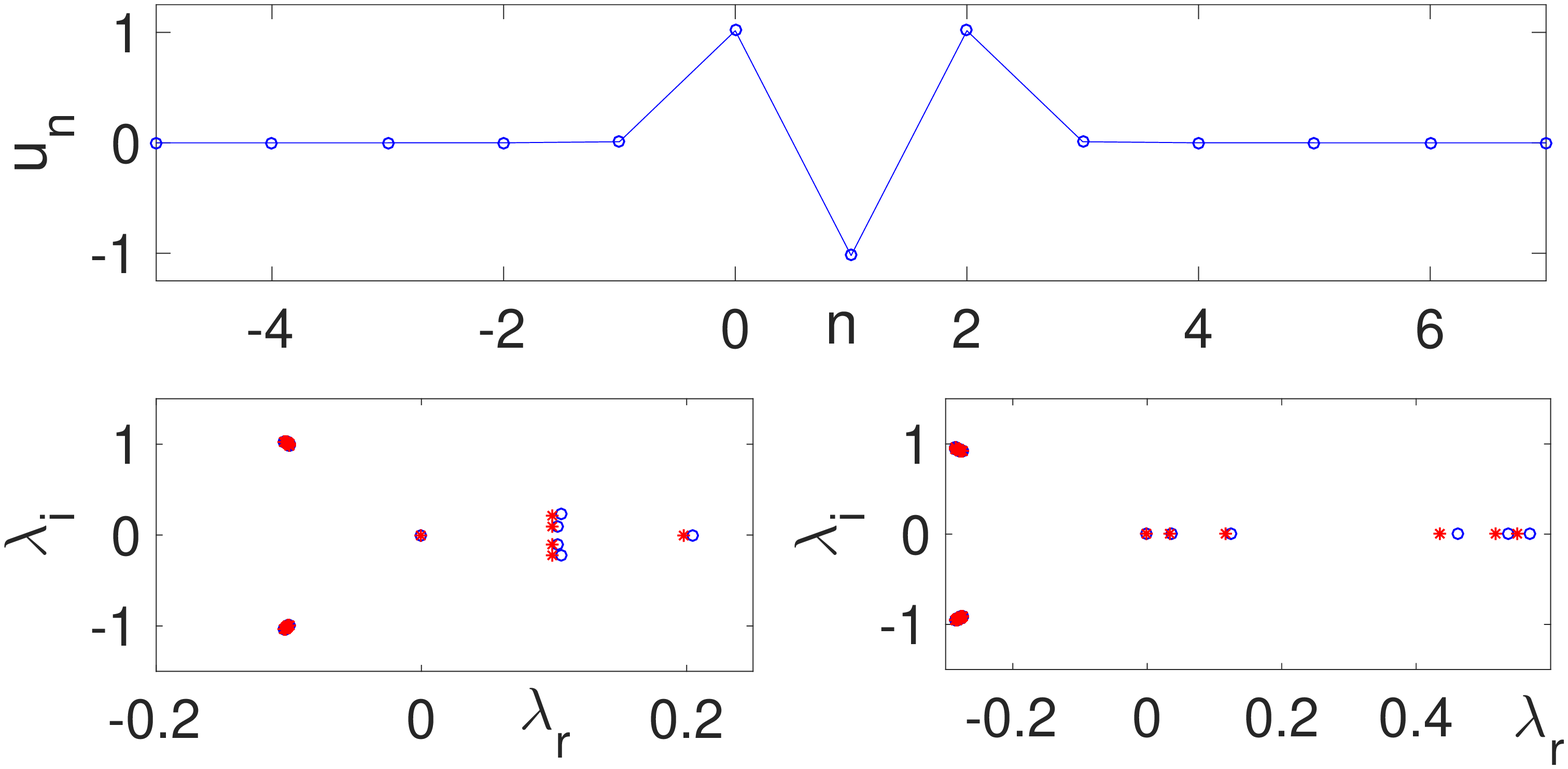}
\includegraphics[height=3.5cm,width=5cm]{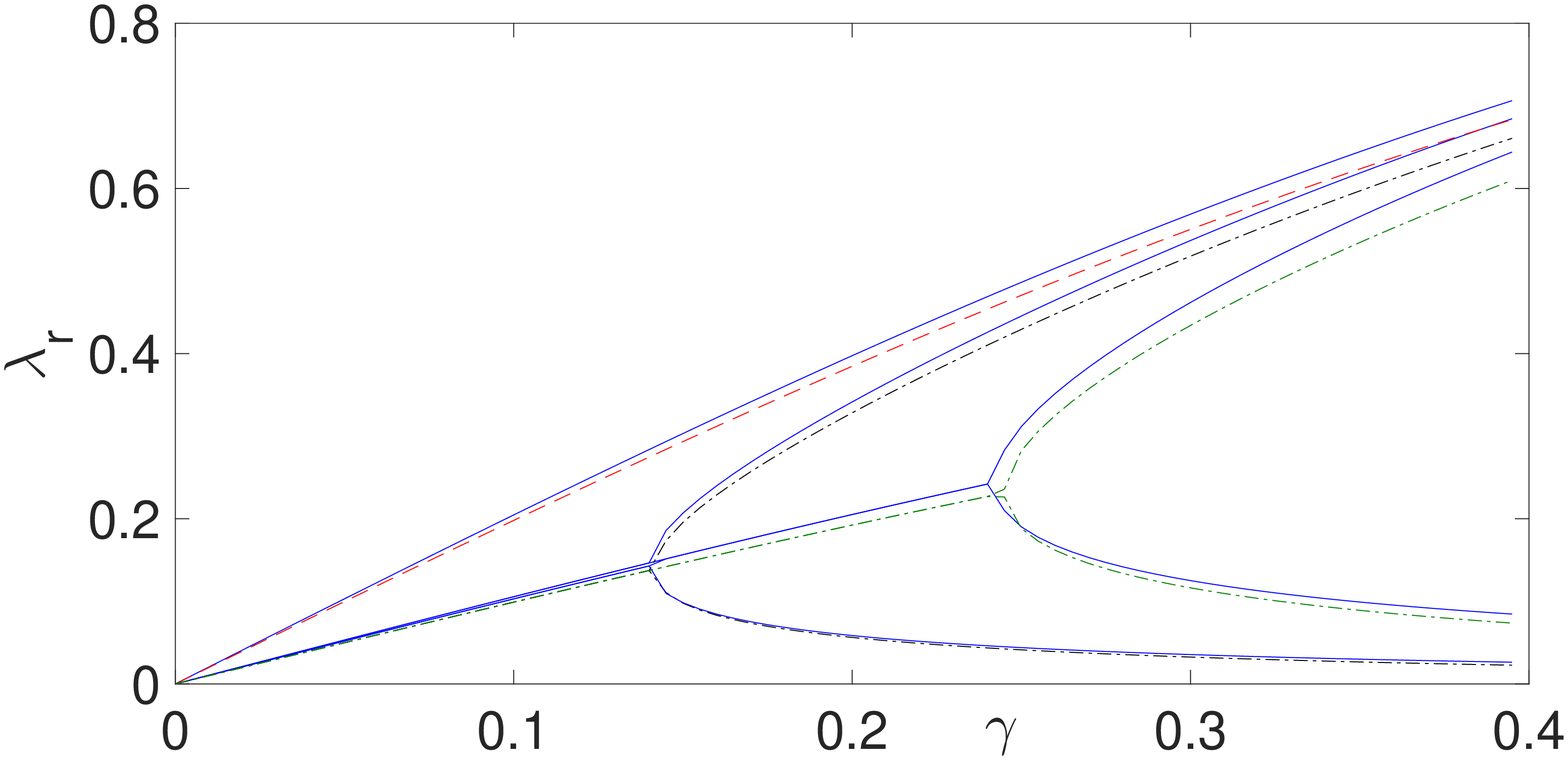}
\includegraphics[height=3.5cm,width=5cm]{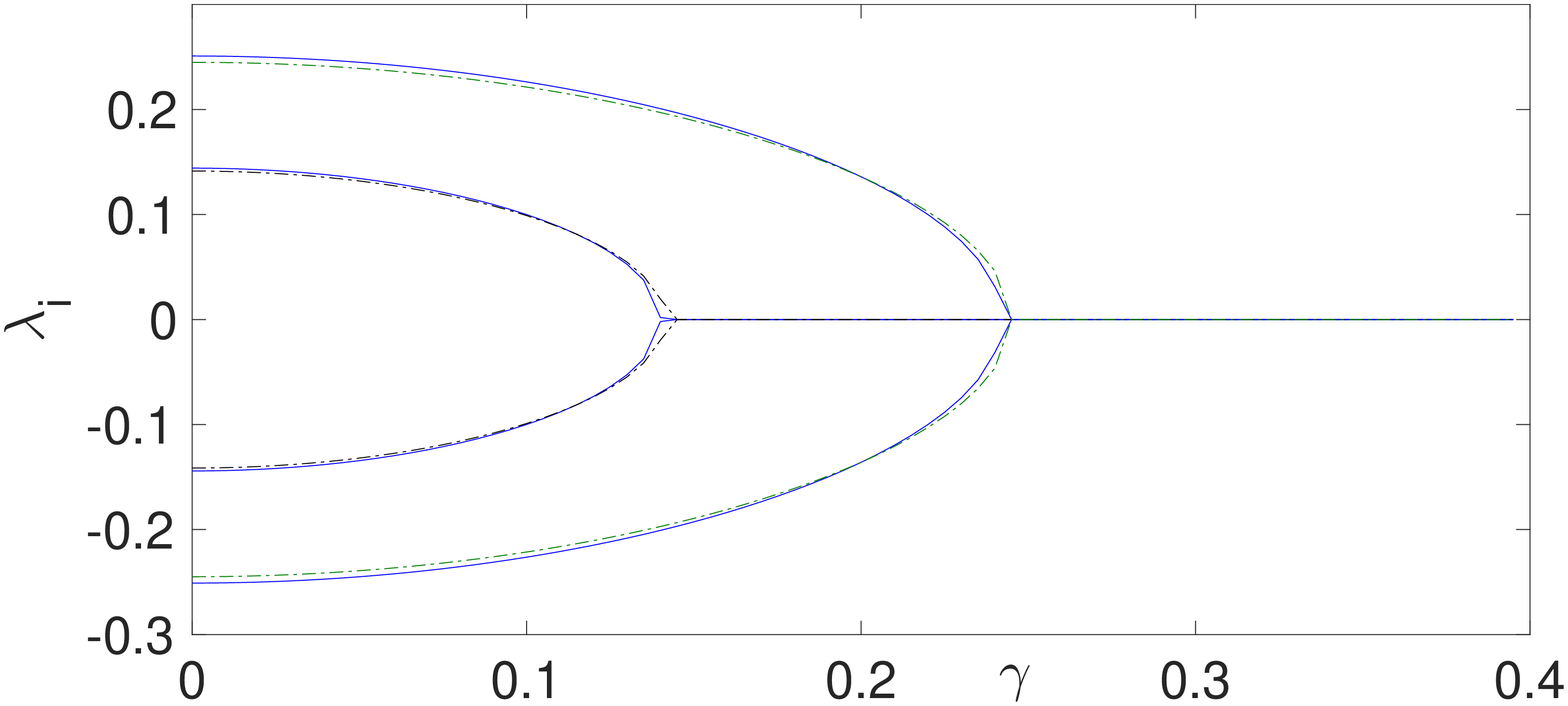}
\caption{Similar to the previous cases (most notably Fig.~\ref{figd2})
  but now for the twisted three-site configuration of opposite
  adjacent signs, i.e., $(+,-,+)$. The left panel shows the
  solution profile and its spectral planes for $\gamma=0.1$
  (with two complex pairs and a real one, as well as the continuous
  spectrum) and for $\gamma=0.3$, past the two critical points,
  where the two complex pairs have collided on the real axis
  becoming real. The configuration, in addition to the (most) unstable
  mode stemming from $\lambda=0$, features the two complex
  --at least for small $\gamma$-- pairs;
  their theoretical prediction is shown in green and black dash-dotted lines
  in the middle and right panel, respectively, for the real and
  imaginary parts. These compare very well with the corresponding numerical
  ones in blue solid lines.
}
\label{figd4}
\end{figure}

Lastly, to illustrate that the relevant phenomenology can, in principle,
be applied to {\it arbitrary} configurations, we consider a
three-site solution. As is well-known, once again from the Hamiltonian
limit, the only one among them that can be stable for finite values
of $\epsilon \neq 0$, is the twisted mode involving three excited
sites and two sign changes between them i.e., $(+,-,+)$ or $(-,+,-)$
if we consider a symbolic representation of the signs of the relevant
solution's excited sites. This configuration is shown in the left panel
of Fig.~\ref{figd4}, along with the corresponding spectral planes for
both $\gamma=0.1$ and $\gamma=0.3$. As the middle panel involving
the real parts and the right panel involving the imaginary ones illustrate,
for this solution, there are two modes that become complex; once again
these stem from the two negative Krein signature modes that are known
to exist in the Hamiltonian limit of $\gamma=0$~\cite{pkf05}.
For these modes, the Hamiltonian limit yields $s=-\epsilon$
and $s=-3 \epsilon$, which implies that they will respectively
be complex --in line with the approximate Eq.~(\ref{dg10})--
until $\gamma_{cr}^{(1)}=\sqrt{2 \epsilon} \approx 0.1414$
and $\gamma_{cr}^{(2)}=\sqrt{6 \epsilon} \approx 0.245$.
The middle and right panels of Fig.~\ref{figd4} indicate
that these critical points, but also the overall behavior of
the relevant eigenvalues is in very good agreement with
the numerical observations. Indeed, prior to the critical points,
the respective pairs are complex, subsequently splitting along the real
axis with two of the four eigenvalues growing and two shrinking
towards $0$. In addition to these modes, there is the single
eigenvalue $2 \gamma/(\gamma^2+1)$, growing from the origin, which
in fact turns out to be the most unstable mode of the system.

%%%%%%%%%%%%%%%%%%%%%%%%%%%%%%%%%%%%%%%%%%%%%%%%%%%%%%%%%%%%%%%%%%%%%%%%%
%%%%%%%%%%%%%%%%%%%%%%%%%%%%%%%%%%%%%%%%%%%%%%%%%%%%%%%%%%%%%%%%%%%%%%%%%

\begin{figure}
\includegraphics[height=4.5cm]{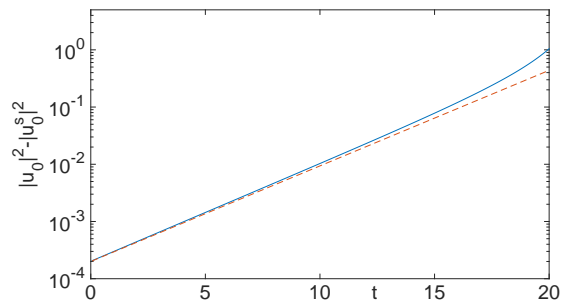}
\caption{For a small initial perturbation to the
  single site solution, we explore how its central
  site $|u_0|^2$ grows in comparison to the steady
  state value $|u_0^s|^2$. The solid blue line
  corresponds to the numerical computation while the
  dashed red line to the theoretical prediction for this growth.
}
\label{figd5}
\end{figure}

We now complement these results with some direct numerical
simulations, to showcase how these findings manifest themselves
therein. We start from a simulation of the single-site solution
illustrated in Fig.~\ref{figd5}. There, we can see that indeed
in this case, as expected from the theory, exponential growth
manifests itself. This is shown by comparing the evolution
of a slightly perturbed central site $|u_0|^2$ with the
corresponding steady state value $|u_0^s|^2$ and seeing
how the difference starting from around $10^{-4}$ grows in
an exponential fashion (blue solid line), linear in the semilog
plot, matching very closely the theoretical dashed
red line, pertaining to the growth rate of the mode
emerging from $(0,0)$ in the spectral plane [recall that its
  growth rate is given by $2 \gamma/(\gamma^2+1)$].

%%%%%%%%%%%%%%%%%%%%%%%%%%%%%%%%%%%%%%%%%%%%%%%%%%%%%%%%%%%%%%%%%%%%%%%%%
%%%%%%%%%%%%%%%%%%%%%%%%%%%%%%%%%%%%%%%%%%%%%%%%%%%%%%%%%%%%%%%%%%%%%%%%%

\begin{figure}
\includegraphics[height=3.5cm,width=5cm]{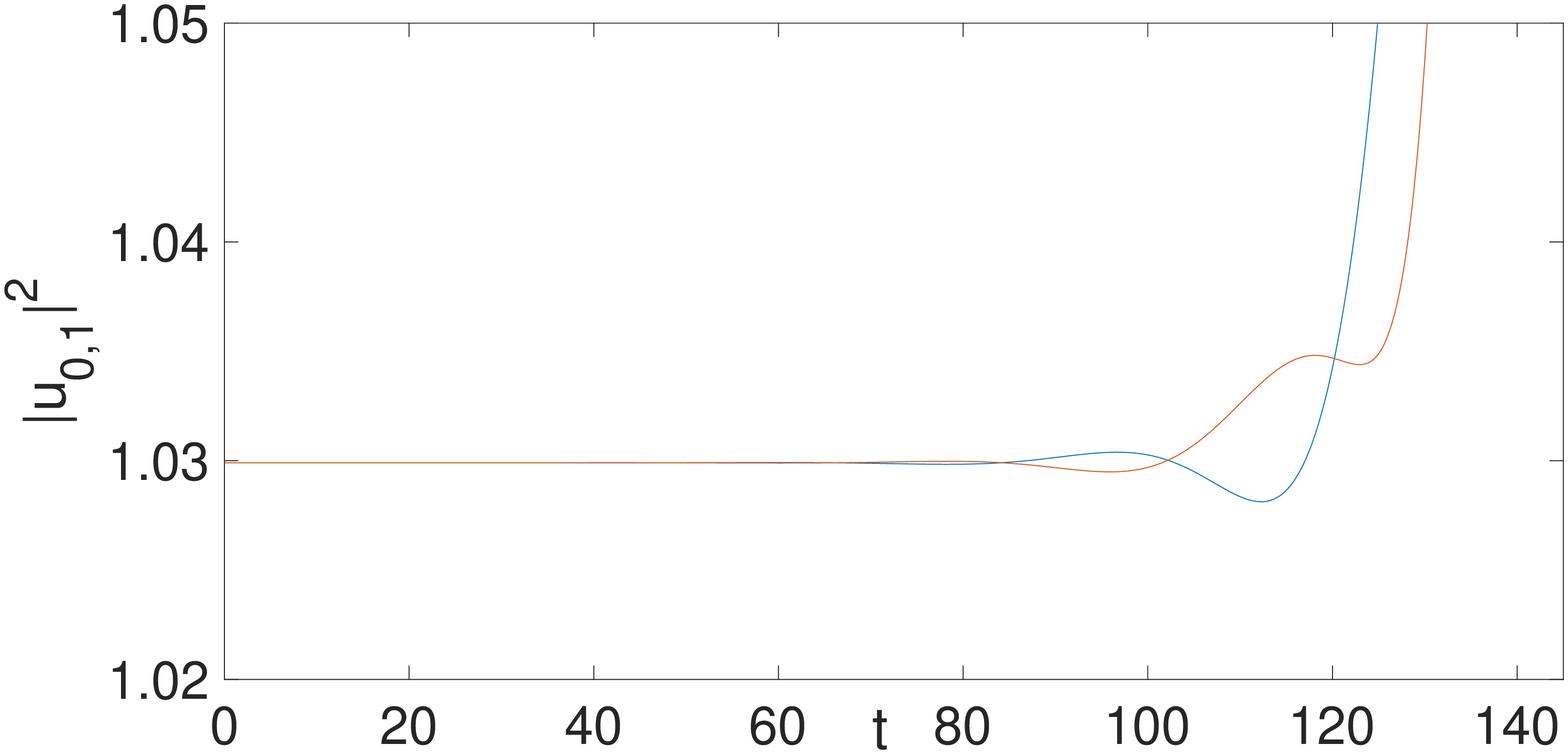}
\includegraphics[height=3.5cm,width=5cm]{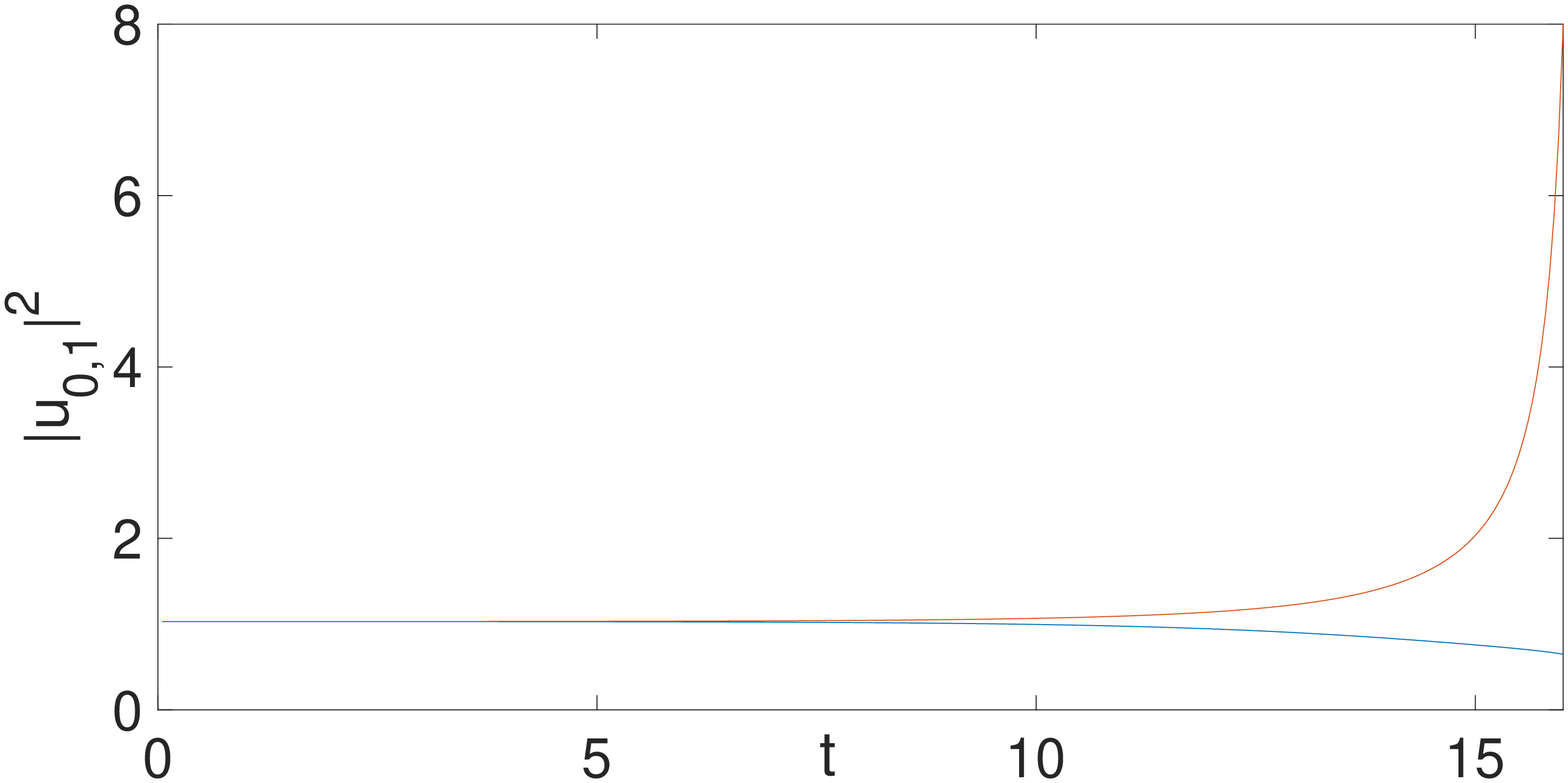}
\caption{Evolution of the two central sites of the solution
  with $n=0$ and $n=1$ in the case of a two-site (twisted mode)
  solution. The left panel featuring oscillatory growth away
  from the steady state is shown for the case of $\gamma=0.1$
  (where a complex pair exists in the linearization), while the right
  panel featuring exponential deviation from the stationary state
  is for the case of $\gamma=0.3$.
}
\label{figd6}
\end{figure}

We next consider the case example of the twisted mode for the
two cases that were already shown in Fig.~\ref{figd2}, namely
$\gamma=0.1$ and $\gamma=0.3$. Fig.~\ref{figd6}
illustrates that in the former case, a fundamentally distinct
feature emerges, namely the emergence of oscillatory growth.
This is natural to expect given the presence of complex eigenvalues
in the system. However, given that they do not constitute the
dominant mode and the growth is fairly rapid, we have suitably
zoomed in to observe how the dynamics concurrently features
oscillations and growth in the early stages of the manifestation
of the instability. This is to be contrasted with the case of
the right panel for $\gamma=0.3$, where the theory predicts
purely real eigenvalues (due to
the complex pairs' collision) and purely exponential growth.
Indeed, in the case shown one of the sites grows while the other
one decays.

%%%%%%%%%%%%%%%%%%%%%%%%%%%%%%%%%%%%%%%%%%%%%%%%%%%%%%%%%%%%%%%%%%%%%%%%%
%%%%%%%%%%%%%%%%%%%%%%%%%%%%%%%%%%%%%%%%%%%%%%%%%%%%%%%%%%%%%%%%%%%%%%%%%

\begin{figure}
\includegraphics[height=4.5cm]{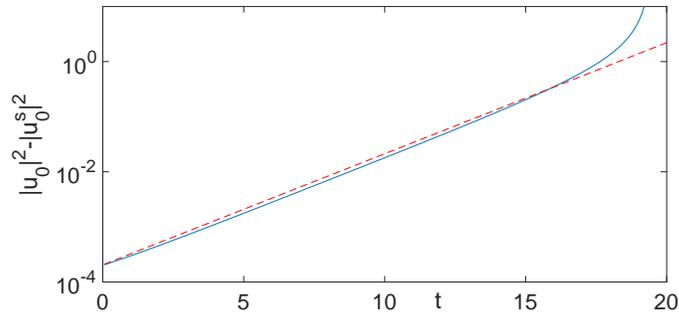}
\caption{Evolution of the $n=0$ node in a two-site in-phase
  solution numerically (blue solid line) vs. the theory of
  the most unstable mode for $\gamma=0.2$ in Fig.~\ref{figd3}.
  Good agreement is again found between the theoretical result
  and the numerical observation.
}
\label{figd7}
\end{figure}

As a final example~\footnote{For the case of three sites, we
  generally observed similar features to the ones above, with
  oscillatory growth dynamics for small enough $\gamma$ and
  exponential growth for sufficiently large $\gamma$ hence
  we do not lend separate consideration to that case here.},
we consider the example of the two-site solution where the
excited sites are in-phase. Fig.~\ref{figd7} captures
the relevant phenomenology illustrating the growth of
one of the two sites of the solution. It can be seen
again that the early stages of the growth are captured by
the relevant mode of the linearization (cf. the most dominant
mode, the one starting as real already for $\gamma=0$
in the case of Fig.~\ref{figd3}). It is worth noting, however,
that although the 2nd (and neighboring) node also features growth
at the early stages, eventually the solution's amplitude symmetry
is broken and one of the two nodes (here the $n=0$ node) grows
faster than the other one.

%%%%%%%%%%%%%%%%%%%%%%%%%%%%%%%%%%%%%%%%%%%%%%%%%%%%%%%%%%%%%%%%%%%%%%%%%
%%%%%%%%%%%%%%%%%%%%%%%%%%%%%%%%%%%%%%%%%%%%%%%%%%%%%%%%%%%%%%%%%%%%%%%%%

%%%%%%%%%%%%%%%%%%%%%%%%%%%%%%%%%%%%%%%%%%%%%%%%%%%%%%%%%%%%%%%%%%%%%%%%%
%\begin{figure}
%\includegraphics[height=4.5cm]{s3_g005_dyn_fig4.eps}
%\includegraphics[height=4.5cm]{s3_fig6_dyn_g03.eps}
%%\includegraphics[height=4.5cm]{s2_fig6_dyn_g03.eps}
%\caption{
%}
%\label{figd8}
%\end{figure}
%%%%%%%%%%%%%%%%%%%%%%%%%%%%%%%%%%%%%%%%%%%%%%%%%%%%%%%%%%%%%%%%%%%%%%%%%

\section{Conclusions \& Future Challenges}

In the present work, we have revisited the widely relevant
DNLS model in the context of a phenomenological non-conservative
term introduced originally for finite temperature
atomic condensates. This
gave us the opportunity to examine
the spectral properties of solitary wave solutions
of the model in the presence of non-conservative perturbations.
%It is interesting to point out here that we have been
%especially careful in our phrasing of this term as being
%a gain/loss one (rather than a purely loss, as it is in
%self-defocusing or repulsive interaction setting). This is
%because, there is
The relevant setting induces a lossy effect at the linear level,
which is responsible for the motion to the left of the
continuous spectrum, however, there is concurrently a gain
effect at the nonlinear level due to the self-focusing nonlinearity.
This introduces an interesting competition in the model.
What we were able to analytically illustrate in the present
setting is that the eigenvalues bifurcating from the origin
(and associated with the nonlinear states of the model)
move in the {\it opposite direction} than the rest of
the spectrum and towards the right half of the spectral
plane, giving rise to instabilities. One such mode comes
from the origin (and is {\it generically} present), while
additional ones stem either from real modes (when present)
and asymmetrize them, or from imaginary ones (when present),
rendering the latter complex, until they collide on the
real axis, at a finite predictable value of the gain/loss
parameter. Remarkably, we saw that the direct analysis
of this non-conservative model eigenvalue problem, combined with a
detailed understanding of the Hamiltonian version of the
relevant operators (borrowed from~\cite{pkf05}) provides
a complete and highly accurate spectral picture for this
system. Lastly, it should be pointed out that the relevant
features (instabilities, via oscillatorily or exponentially
growing modes) have also been corroborated via direct numerical
simulations.

It would be interesting to extend the present considerations
to different classes of systems.
%As indicated above,
%the presence of the $\gamma$ term on the left hand side
%(the time derivative) features a complex and nontrivial
%form of including both loss and gain. However,
One can
consider other forms involving gain and loss in NLS
(see, e.g., for a relevant recent example~\cite{nikoshir}),
hence it would be relevant to explore to what degree
the considerations herein would apply there.
In particular, the analysis presented here does not
address the general dissipative case; it is more restricted
to (weak) perturbations of the Hamiltonian problem that
destroy its Hamiltonian nature.
Nevertheless, we expect
that some of these properties (e.g., the motion of eigenvalues
of negative Krein signature under non-conservative
perturbations) should be of broader relevance than the
specific context considered herein.
However, exploring the extent to which these topological considerations
apply and where they may fail will be of particular interest for
future studies.
It would also
be of interest to explore this, as well as other
models in higher dimensional settings to examine their
implications on both soliton and the more elaborate
vortex solutions featured in the latter~\cite{dnlsbook}.
Such studies will be considered in future publications.

\end{document}